\documentclass[conference,9pt]{IEEEtran}
\IEEEoverridecommandlockouts

\usepackage{cite}
\usepackage{amsmath,amssymb,amsfonts}
\usepackage{algorithmic}
\usepackage{graphicx}
\usepackage{svg} 
\usepackage{textcomp}
\usepackage[dvipsnames]{xcolor}
\def\BibTeX{{\rm B\kern-.05em{\sc i\kern-.025em b}\kern-.08em
    T\kern-.1667em\lower.7ex\hbox{E}\kern-.125emX}}

\usepackage[acronym,nomain]{glossaries}
\usepackage[free-standing-units,per-mode=repeated-symbol,detect-weight=true,detect-family=true,retain-explicit-plus=true]{siunitx}

\usepackage{pgfplots}
\pgfplotsset{compat=1.18}
\usepackage{pgfplotstable}
\usepackage{booktabs}
\usepackage{listings}

\DeclareSIUnit{\unitless}{\relax}

\DeclareSIUnit\bits{bits}
\DeclareSIUnit\bps{bps}
\DeclareSIUnit\Bps{Bps}
\DeclareSIUnit\request{request}
\DeclareSIUnit\cycle{cycle}
\DeclareSIUnit\flop{FLOP}
\DeclareSIUnit\flops{FLOPS}
\DeclareSIUnit\gate{GE}
\DeclareSIUnit\op{OP}
\DeclareSIUnit\ops{OPS}
\DeclareSIUnit\ipc{IPC}

\definecolor{color_technology}{HTML}{000000}
\definecolor{color_architecture}{HTML}{000000}
\definecolor{color_reliability}{HTML}{000000}
\definecolor{color_eda}{HTML}{000000}

\newcommand\suppress[1]{}

\let\OldTexttrademark\texttrademark
\renewcommand{\texttrademark}{\OldTexttrademark\ }
\newacronym{PULP}{PULP}{Parallel Ultra Low Power}
\newacronym{EDA}{EDA}{Electronic Design Automation}
\newacronym{DTCO}{DTCO}{Design Technology Co-Optimization}
\newacronym{STCO}{STCO}{System Technology Co-Optimization}
\newacronym{PPA}{PPA}{Power-Performance-Area}
\newacronym{PDN}{PDN}{Power Distribution Network}
\newacronym{CFET}{CFET}{Complementary FET}
\newacronym{VLSI}{VLSI}{Very-large-scale integration}
\newacronym{BSPDN}{BSPDN}{Backside Power Distribution Network}
\newacronym{PPACE}{PPACE}{Power-Performance-Area-Cost-Environment}
\newacronym{CSS}{CSS}{Compute Subsystems}
\newacronym[shortplural={FEOLs}]{FEOL}{FEOL}{Front End Of Line}
\newacronym[shortplural={BEOLs}]{BEOL}{BEOL}{Back End Of Line}
\newacronym{LD}{LD}{Low-Density}
\newacronym{HD}{HD}{High-Density}
\newacronym{LP}{LP}{Low-Power}
\newacronym{HP}{HP}{High-Performance}
\newacronym{F2F}{F2F}{Face-to-Face}
\newacronym{F2B}{F2B}{Face-to-Back}
\newacronym{W2W}{W2W}{Wafer-to-Wafer}
\newacronym{D2W}{D2W}{Die-to-Wafer}
\newacronym{HPWL}{HPWL}{Half-Perimeter Wirelength}
\newacronym{SLM}{SLM}{Silicon Lifecycle Management}
\newacronym{DFT}{DFT}{Design For Test}
\newacronym{P-PDK}{P-PDK}{Pathfinding-PDK}
\newacronym{TSV}{TSV}{Through-silicon-via}
\newacronym{LOL}{LOL}{Logic-on-Logic}
\newacronym{MOL}{MOL}{Memory-on-Logic}
\newacronym{LOM}{LOM}{Logic-on-Memory}
\newacronym{AuC}{AuC}{Array-under-CMOS}

\newacronym{PE}{PE}{Processing Element}
\newacronym{HBM}{HBM}{High Bandwidth Memory}
\newacronym{C2C}{C2C}{Chip-to-Chip}
\newacronym{CGRA}{CGRA}{coarse-grained reconfigurable array}
\newacronym{CIM}{CIM}{Compute-in-memory}

\begin{document}
\bstctlcite{IEEEexample:BSTcontrol}
\makeatletter
\def\@IEEEpubidpullup{3\baselineskip}
\makeatother

\title{Invited Paper: CMOS~2.0 - Redefining the Future of Scaling}

\def\afflimec{\textsuperscript{1}}
\def\affleth{\textsuperscript{2}}
\def\afflimecuk{\textsuperscript{3}}
\def\afflkit{\textsuperscript{4}}
\def\afflbologna{\textsuperscript{5}}

\author{
Moritz Brunion\afflimec,
Navaneeth Kunhi Purayil\affleth, Francesco Dell'Atti\afflimec, Sebastian Lam\afflimecuk, Refik Bilgic\afflimec, \\
Mehdi Tahoori\textsuperscript{4,1}, Luca Benini\textsuperscript{2,5}, Julien Ryckaert\afflimec\\
\\
\textsuperscript{1}\emph{imec, Leuven, Belgium}\quad
\textsuperscript{2}\emph{ETH Zürich, Zürich, Switzerland}\quad
\textsuperscript{3}\emph{imec, Cambridge, UK}\\
\textsuperscript{4}\emph{Karlsruhe Institute of Technology, Karlsruhe, Germany}\quad
\textsuperscript{5}\emph{Università di Bologna, Bologna, Italy}\\
mehdi.tahoori@kit.edu, lbenini@iis.ee.ethz.ch, julien.ryckaert@imec.be
}

\newcommand{\copyrightnoticearxiv}{
    \IEEEoverridecommandlockouts
    \IEEEpubid{
    \parbox{\columnwidth}{\vspace{-4\baselineskip}\copyright 2025 IEEE. Personal use of this material is permitted. Permission from IEEE must be obtained for all other uses, in any current or future media, including reprinting/republishing this material for advertising or promotional purposes, creating new collective works, for resale or redistribution to servers or lists, or reuse of any copyrighted component of this work in other works. DOI: 10.1109/ICCAD66269.2025.11240655
    \hfill}
    \hspace{0.9\columnsep}\makebox[\columnwidth]{\hfill}}
    \IEEEpubidadjcol
}
\copyrightnoticearxiv

\maketitle

\IEEEpubidadjcol

\begin{abstract}
We propose to revisit the functional scaling paradigm by capitalizing on two recent developments in advanced chip manufacturing, namely 3D wafer bonding and backside processing. This approach leads to the proposal of the CMOS~2.0 platform. The main idea is to shift the CMOS roadmap from geometric scaling to fine-grain heterogeneous 3D stacking of specialized active device layers to achieve the ultimate Power-Performance-Area and Cost gains expected from future technology generations. However, the efficient utilization of such a platform requires devising architectures that can optimally map onto this technology, as well as the EDA infrastructure that supports it. We also discuss reliability concerns and eventual mitigation approaches. This paper provides pointers into the major disruptions we expect in the design of systems in CMOS~2.0 moving forward.
\end{abstract}


\section{Introduction}

\begin{figure}
    \centering
    \includegraphics[width=\linewidth]{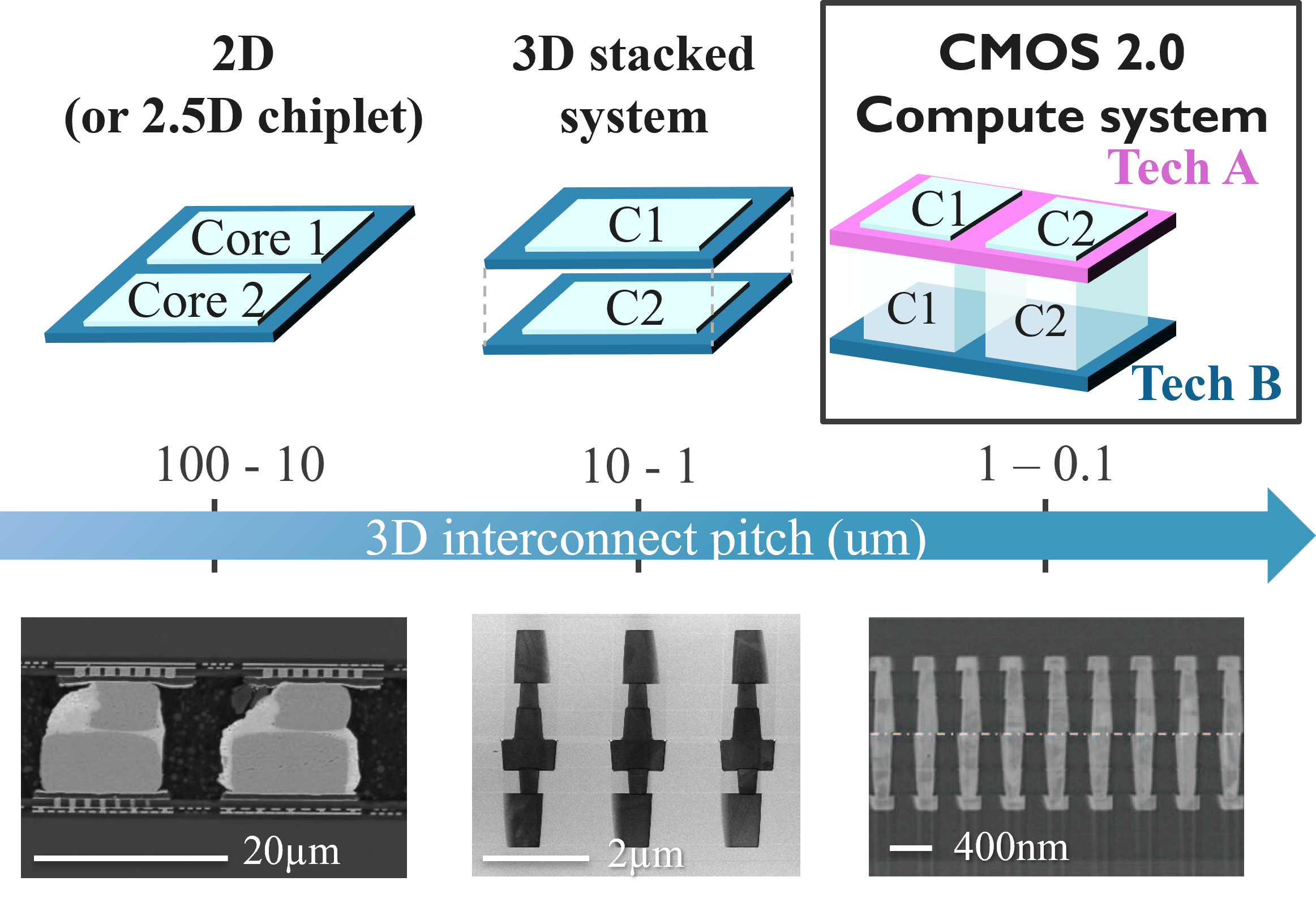}
    \caption{CMOS~2.0 versus 3D-IC}
    \label{fig:technology:CMOS2.0vs3DIC}
\end{figure}

As the semiconductor industry advances into the Angstrom era, geometric scaling becomes extremely challenging. Research efforts are increasingly focused on the question: What comes after \gls{CFET}~\cite{cfet2018}? A relief from the staggering challenges of pure geometric scaling comes from the recent promising advancements in fine pitch 3D bonding \cite{chew2024_400nm_pitch}.\gls{W2W} hybrid bonding, as shown in Fig.~\ref{fig:technology:CMOS2.0vs3DIC}, now enables two paradigm shifts: sub-micron inter-die connectivity and the stacking of heterogeneous \gls{FEOL} layers with specialized \gls{BEOL}. This opens the path to technology-aware system design exploiting a heterogeneous third dimension.

The \gls{PPA} gains in this scenario no longer stem exclusively from classical CMOS scaling; we therefore refer to this shift as \textbf{CMOS~2.0}. It represents a major expansion of the CMOS design space, introducing richer characteristics, broader design freedoms, and tighter cross-layer interactions (Fig.~\ref{fig:technology:CMOS2.0vs3DIC}). This perspective emerges once the design problem is freed from the constraints of planarity and the homogeneity of a single \gls{FEOL} technology.\\
Long-standing assumptions in \gls{VLSI} must consequently be revisited. Because this problem spans multiple domains, this paper articulates key questions raised by this transition. It develops them, not to provide definitive answers, but to catalyze the community to identify root drivers and respond with timely, principled solutions to a broad challenge. We foresee it to be instrumental for future technological advancements.

First, we examine the physical scaling walls faced by current technologies and the bonding granularities that can practically enable CMOS~2.0. Second, we analyze architectural constraints inherent to 2D integration and theorize the design benefits of exploiting a third dimension with heterogeneous integration. Third, we outline significant implementation challenges that \gls{EDA} tools must address to fully leverage the technology through more informed mapping. Fourth, we discuss reliability implications and possible mitigation approaches to such dense heterogeneous 3D systems.
\section{Towards CMOS~2.0}
\label{sec:Technology}

\color{color_technology}

\subsection{The CMOS scaling challenge}
Ever since Moore's Law set the course of digital scaling, its pace has essentially been dictated by geometry reduction. The objective was straightforward: the scaling of the CMOS switch and the interconnect fabric around it was done such that the functional density scaling it entails improves the overall power and performance of the system. How scaling density ended up entangled with power and performance improvement has, however, evolved. From Dennard scaling to the exploitation of multi-core architectures and, more recently, through highly specialized cores and accelerators, the ultimate objective was to find a complementarity between geometric scaling and performance/power improvements. The success and rapid deployment of systems under this geometric reduction principle were, for a big part, due to the simplicity of its evaluation and implementation. Indeed, the continuity in the geometric scaling of CMOS allowed the development of a layered ecosystem of specialized tasks all connected through a complex set of abstraction levels. From device-level TCAD optimization to the design automation of billions of objects, each layer could define a set of guiding principles to model, calibrate, and ultimately drive the scaling trajectory in a concurrent manner. A second yet important reason for the success of CMOS lies in its general-purpose nature. By its capability to emulate diverse functions, it enabled the System-on-Chip paradigm, where reducing off-chip data movement significantly boosts energy efficiency. Modern systems now integrate a large set of diverse processing elements on the same substrate, each leveraging the benefits of CMOS scaling.

Moving forward, scaling will become increasingly challenging, potentially affecting the foundations of this entire ecosystem. At its heart is the issue of defining a CMOS switch that satisfies the Power-Performance-Area (PPA) scaling constraints. This problem is not new and is at the origin of the Design-Technology Co-Optimization (DTCO) Era, coinciding with the development of finFET nodes. On top of mitigating short-channel effects by confining the Si channel, finFET devices have allowed a series of structural modifications at the standard cell level that maximize the amount of driving current from a given area. Complemented with DTCO boosters such as Single-Diffusion Break (SDB) or Contact-Over-Active-Gate (COAG), standard cell co-optimization with the device architecture has become the engine that was used to produce a series of successful technology nodes since 20nm. The more recent Gate-All-Around (GAA) device, in the direct continuity of this evolution, maximizes current drive over device parasitic cap under a scaling constraint. Down the road, the CFET device following GAA will likely be the ultimate scaled Si FET architecture. Built on a stacked configuration of nFET and pFET, it attempts to remove all possible area overhead in a CMOS standard cell. Beyond CFET, there is currently no device architecture proposal that seems to have the capability to sustain the scaling trajectory. Certainly, CNT and 2D-material-based devices could provide a path forward, but their ability to maintain PPA scaling as set by the Si Era is still to be proven. Bottom line, in the absence of a universal scaled switch, we need to find other scaling mechanisms for technology to support the ever-increasing demand for more efficient compute performance. This is where CMOS~2.0 can offer a path forward.

\subsection{The premise of CMOS~2.0}

\begin{figure}
    \centering
    \includegraphics[width=\linewidth]{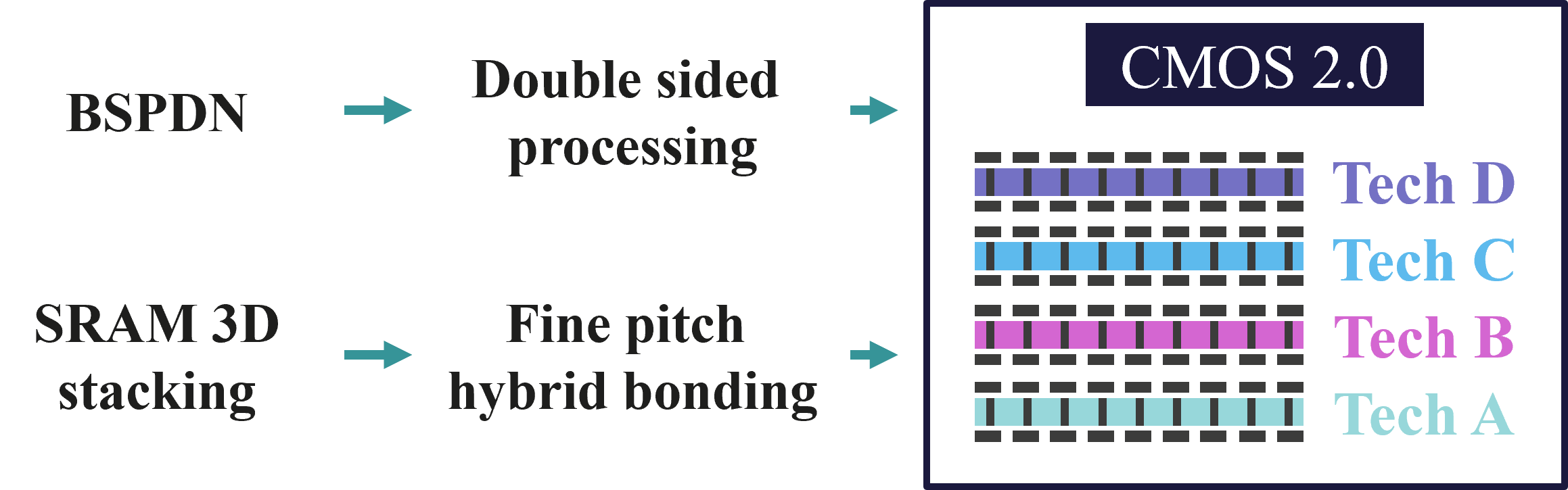}
    \caption{CMOS~2.0 emerging from hybrid bonding and double-sided processing}
    \label{fig:technology:CMOS2.0origin}
\end{figure}

Beyond device scaling, the demand for more compute performance has also pushed technology manufacturers to explore other vectors for scaling systems. Among these innovations, two are worth mentioning as they are foundational to the development of CMOS~2.0 (Fig. \ref{fig:technology:CMOS2.0origin}). 

The first one is 3D assembly technology. Driven by the necessity to increase on-chip memory in processor chips and the need to scale DRAM technologies, die-to-wafer and wafer-to-wafer assembly have recently migrated to hybrid bonding techniques. This process has enabled a drastic downscaling of bonding pitches in the sub-$\mu$m range, with a record of \SI{400}{\nano\meter} as shown in \cite{zhang2024_400nm_pitch}. Achieving such a granularity of connections between active device tiers challenges the SoC paradigm itself, as connecting objects between different substrates does not necessarily induce energy inefficiencies anymore. Even better, each technology tier can now be optimized for the function itself, potentially minimizing the overall PPA and Cost of the System.

The second system-level technology innovation is the double-sided processing of CMOS wafers. Driven by the backside power delivery network available in sub-2nm nodes, the enrichment of backside connectivity will soon allow device-level terminals to be exposed on the backside. In some CFET integration propositions, the device itself could even potentially be partially completed from the backside. This will allow future technologies to exhibit symmetric access to the front- and back-side of the die, with devices being sandwiched between two fine-grain interconnect layers.
The vision of a CMOS~2.0 technology platform consists of a stack of active tiers all connected through a fine-grain interconnect network, where each device layer is tailored to a specific function it would serve in the system. This vision extends the more classic 3D-IC stacking approach towards a co-design of a functional entity across tiers as shown in Fig.~\ref{fig:technology:CMOS2.0vs3DIC}. Capitalizing on fine-grain hybrid bonding and double-sided processing, the aim of CMOS~2.0 is to offer a platform composed of heterogeneous tiers, enabling flexible and efficient system mapping, as illustrated in Fig. \ref{fig:technology:CMOS2.0origin}.

With geometric scaling no longer sustaining CMOS advancement, CMOS~2.0 introduces a paradigm shift that enables continued progress through heterogeneous 3D integration. Imposing a unified device and interconnect stack to support all digital system functions constrains the technology in ways that struggle to meet the heterogeneous requirements of SoC architectures. On a global scale, data parallelism has generalized the usage of multiple processing cores and specific accelerators. The PPA optimization for each of these processing units individually often comes into conflict with a general-purpose technology platform. Also, SoCs feature complex system caches as well as a high-speed connectivity fabric. Both of these would benefit from very different device and interconnect technology settings. Within a logic block itself, one can also distinguish a very diverse set of operations. Between sequential logic gates and combinational logic gates, very different technology trade-offs could be made if taken individually. Even within the combinational network, we can distinguish dense local logic networks from global networks, ideally requiring vastly different interconnect and device settings. For example, a choice in the number of Mx layers as the lowest, aggressively scaled metal layers in the BEOL is mainly driven by the necessity to support local dense routing. A high number of Mx layers, in turn, penalizes other parts of the system operating data on a more global scale, with an increase in interconnect resistance when accessing less resistive My and Mz layers on top of the Mx layers. The device itself today needs to be dimensioned for timing critical paths, setting the device width to an unnecessarily large value when it comes to the majority of the other logic gates. Oversized device width necessitates multiple stacked sheets, increasing self-loading capacitance and leading to power inefficiencies in dense logic regions. Finally, the power and the clock network require a vastly different set of devices and interconnects than what is usually needed in dense logic.

As CMOS~2.0 partitions the system into a dense heterogeneous 3D interconnected stack, its technical as well as economic success will depend on the ability to map various compute architectures on a stack that maintains a general-purpose property. This poses four fundamental questions that we will try to address in this paper:

\begin{enumerate}
    \item \label{tech:teaser_tech} \emph{Which partitions can CMOS~2.0 stacks be composed of?}
    \item \label{tech:teaser_arch} \emph{How to design compute architectures for CMOS 2.0?}
    \item \label{tech:teaser_eda} \emph{How to automate placement and routing in such stacks?}
    \item \label{tech:teaser_reliability} \emph{How to address reliability concerns in CMOS~2.0?}
\end{enumerate}

Item \ref{tech:teaser_tech}) will be discussed in subsection \ref{techno_platform}, item \ref{tech:teaser_arch}) will be the subject of Section \ref{sec:architecture}, item \ref{tech:teaser_eda}) will get addressed by Section \ref{sec:EDA} while item \ref{tech:teaser_reliability}) will be discussed in Section \ref{sec:reliability}.

\subsection{A functionality-driven CMOS~2.0 technology stack}
\label{techno_platform}

The name \emph{CMOS 2.0} was chosen deliberately. As discussed earlier, the success of CMOS scaling came from the fact that it enabled a general-purpose roadmap, allowing a sophisticated IP and design ecosystem. Attempting to disrupt it may prove economically unviable. Instead, we need to find a way to leverage this CMOS ecosystem and build from its most efficient property, which is to enable compact power-efficient Boolean logic gates. In this paper, we propose a CMOS~2.0 partitioning strategy that breaks any digital physical implementation at its root, without disrupting standard CMOS logic functional operations. As digital compute units process information coded in discrete bit streams, we can recognize four key operations.

\begin{itemize}
    \item \emph{Transform}: Combinational operations
    \item \emph{Sample and hold}: Sequencing logic
    \item \emph{Store}: SRAM, Register Files,...
    \item \emph{Distribute}: NoC, Bus,...
\end{itemize}

\begin{figure}
    \centering
    \includegraphics[width=\linewidth]{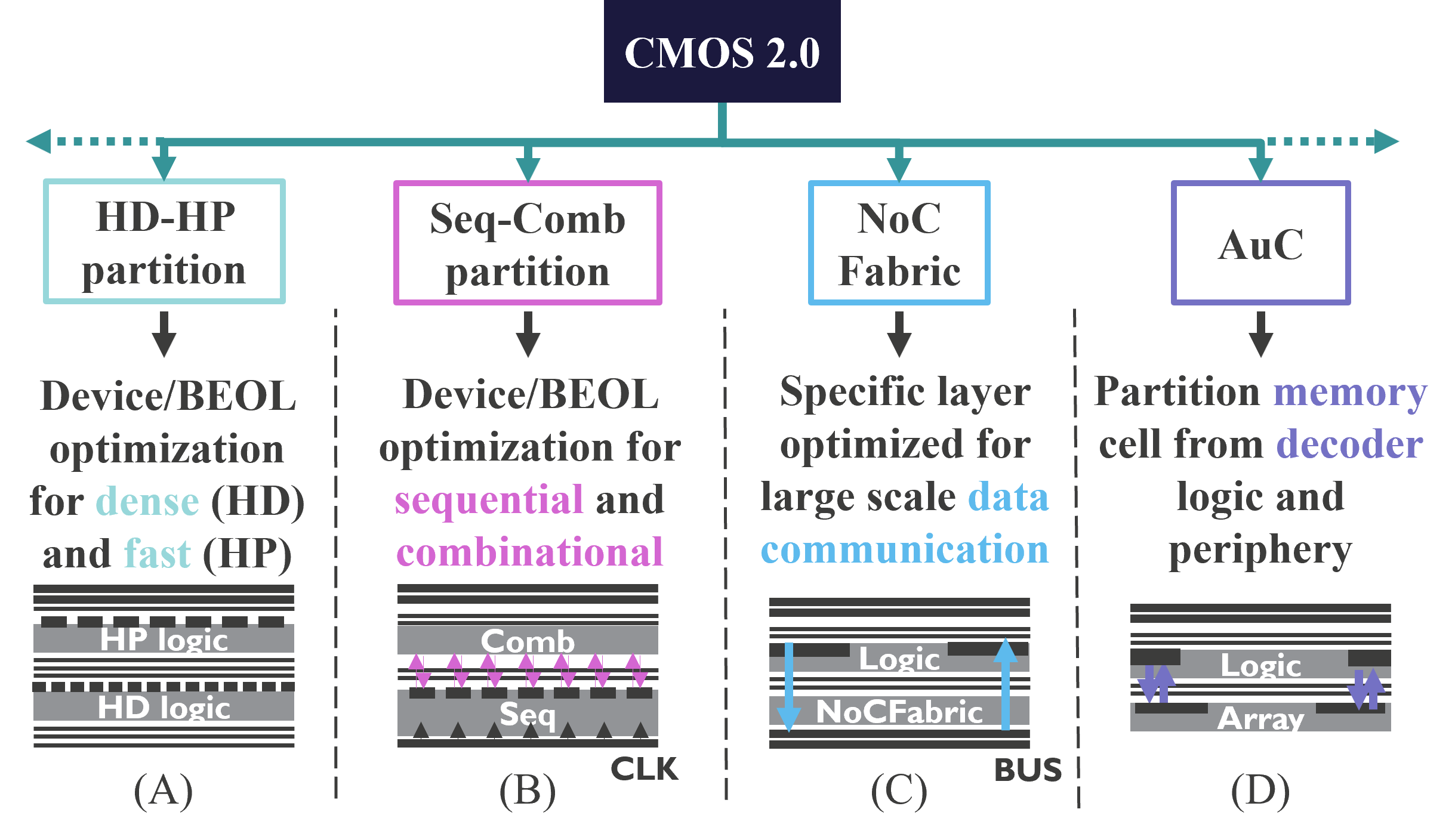}
    \caption{Four potential fine-grain partitioning in digital systems}
    \label{fig:technology:CMOS2.0partition}
\end{figure}

Each of these operations requires devices that feature different types of properties. Various partitioning schemes can then be envisioned using the CMOS~2.0 approach as listed in Fig.~\ref{fig:technology:CMOS2.0partition}. Combinational logic gates need an extensive menu of compact CMOS functions. Being the most significant fraction of the overall design, achieving compact, low-power arrangements of these objects is key to scaling. They will drive tiny low-capacitance devices and ultra-dense interconnect around them. However, a large fanout or BEOL interconnect load will limit the speed of some of the data paths, usually requiring higher current drive. These speed paths set a performance constraint for the intrinsic CMOS device, unnecessarily increasing the total pin cap of the design. If we can partition a dense energy-efficient logic from a faster high-drive layer (case A), we can potentially build way more energy-efficient implementations. A second partitioning scheme separates the combinational from sequencing logic (case B). The latter maintain reliable operations by building synchronicity. They rely on a latching operation, are usually quite large in area, and are assembled in multi-bit structures. It is interesting to note that in a classic register-to-register data path, each register connects to combinational cells. A partitioning at that level would lead to a mostly out-of-plane data flow (tier-to-tier). This relaxes the interconnect requirements for the metal stack belonging to the flip-flop layer.

At a larger scale, SoCs featuring multiple processing units, IO circuits, and storage elements require Buses and Networks-on-Chip (NoC) that allow the high-speed distribution of data across the system. Such an infrastructure requires large buffers and low resistance lines. Using the scaled device of advanced nodes is not effective in this context, and a specific technology could be envisioned for this particular layer (case C).

Finally, the storage of information is distributed across various layers of the compute architecture. From register files and L0 caches inside the processing core up to shared cache resources between processing units, they have different capacity and speed requirements and hence utilize various types of circuitry. A 3D partitioning of the highest cache levels is already implemented in several products today. Going deeper into the lower-level memory structures, critical speed requirements usually limit their capacity. In a fourth partitioning scheme, we propose to disintegrate the memory array from part of its periphery. Encoding/decoding logic can still be implemented in the advanced logic tier, but the bitcells could drive a completely different technology optimization (case D). This partitioning scheme is being studied in some memory technologies and referred to \gls{AuC} \cite{auc_shairfe}.

Each of these four partitioning schemes induces various degrees of microarchitecture modification, as will be discussed in Section \ref{sec:architecture}. Moreover, their complement allows the implementation of a composite heterogeneous stack that combines several of these partitioning schemes together. In this way, CMOS~2.0, although maintaining a large part of the CMOS ecosystem, can vastly affect the physical implementation of digital architectures. It can drive new architectures that map more optimally to such a hierarchical 3D stack, require new EDA that can leverage its full potential, and even allow tailoring each technology layer. Nevertheless, there are several constraints imposed by CMOS~2.0. First, the density scaling moving from 2D to 3D can lead to severe thermal and power issues that could potentially limit its scaling potential. It requires specific technology innovations and will be briefly discussed in Section \ref{sec:reliability}. 

A second significant challenge is the balanced utilization of each layer across the stack. Indeed, a CMOS~2.0 stack will most likely have to be implemented as a wafer-to-wafer process for two reasons: W2W offers the tightest pitch capability, and fine-grain breakdown requires distributing resources across the die, as localizing them would undermine the goals of CMOS~2.0. An efficient area utilization will be key to ensuring a cost-effective technology. We will discuss the architecture implications in Section \ref{sec:architecture} and an approach to the design automation in Section \ref{sec:EDA}.

\color{black}
\section{Architecture}
\label{sec:architecture}

\color{color_architecture}

\begin{figure}[!h]
    \centering
    \includegraphics[width=0.9\linewidth]{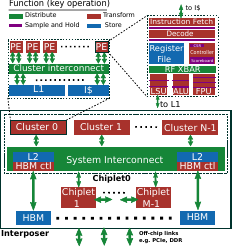}
    \caption{Block diagram of a typical parallel system depicting various functional elements blocks at the chiplet, cluster, and the \gls{PE} level.}
    \label{fig:arch_template}
\end{figure}

CMOS~2.0 requires a profound rethinking of system architecture. Consider a general abstract template for a highly parallel integrated architecture, as shown in Fig.~\ref{fig:arch_template}. The basic building block is a programmable \gls{PE} with fetch, decode, and execution pipelines as well as L0 memories (register files). Several \glspl{PE} together share a local L1 memory, forming a cluster, and multiple clusters are interconnected at the system level to L2/L3 memories and external memory controllers, forming a compute chiplet. At the substrate level, several of these chiplets communicate with off-chip main memory, such as \gls{HBM}, and with other chiplets over \gls{C2C} links. We can assign the components of such an architecture to the four key functions introduced in section \ref{techno_platform}, namely: transform (compute - combinational logic), distribute (communicate - interconnects), sample and hold (pipeline registers, state registers), and store (memories). CMOS~2.0 enables a deep revision in the design of all these components, as well as different ways to distribute and mix them. 

Key general directions are: (i) increasing the amount of local storage, tightly coupled with computing via larger register files, deeper pipelines, near and in-memory computing, (ii) enabling aggressive scale-up of low latency high bandwidth interconnects by exploiting the greatly increased wire budget and dedicating layers to interconnect routing resources, and (iii) creating novel opportunity for reconfigurability at coarse and fine grain. 

In the next subsections, we give concrete examples of architectural innovation opportunities along these directions. We categorize opportunities enabled by CMOS~2.0 into three progressive tiers:
\begin{itemize}
    \item 3D enhancements to existing architectures
    \item Micro-architectural innovations for 3D
    \item Novel architectural paradigms
\end{itemize}
\subsection{3D enhancements to existing architectures}
3D technology can enable substantial performance and energy-efficiency improvements, exploiting the increased spatial freedom and fine pitch die-to-die bonding, allowing tighter integration between compute and memory.
For instance, 3D splitting of SRAMs and better placement flow were demonstrated to achieve higher PPA on an Arm Cortex-A processor \cite{xu_enhanced_2019}. Similarly, hybrid bonding techniques between heterogeneous technology nodes—where, for example, a faster bottom die handles timing-critical paths, while the wiring-critical logic leverages a slower top die for better routing, have been explored \cite{gopireddy_designing_2019}.
In industry, one of the most notable examples of architectural enhancement enabled by three-dimensional heterogeneous integration is AMD's 3D V-Cache technology \cite{amd_3D}. V-Cache is a commercial success. It allows the integration of a 3$\times$ larger L3 cache closer to compute cores compared to classical designs, significantly reducing off-chip memory accesses, achieving higher performance and energy efficiencies.

While the 2-tier design partitioning supported by leading foundries today can offer \gls{PPA} benefits, CMOS~2.0 enables aggressive fine-grain partitioning into several layers with different functional characteristics. This allows for the design of more complex dataflow architectures.
For example, fine-grained partitioning into several combinational and sequential logic tiers could allow for deeper pipelining, boosting the peak operating frequencies.
Furthermore, features like backside power delivery can recover routing resources in the \gls{BEOL} metal stack previously allocated for both power delivery and signal routing.

\subsection{Micro-architectural innovations for 3D}
The increased spatial degrees of freedom available in the third dimension can be exploited to increase on-chip memory size, bandwidth, and reduce interconnect latencies not only as PPA boosters in traditional architectures, but even natively targeting architectures to the new implementation fabric. 
Shared-memory many-PE cluster architectures, such as Mempool and Terapool \cite{mempool,terapool}, feature wire-intensive low-diameter interconnects between PEs and many-bank L1 shared memory. These interconnects are extremely unwieldy to implement within a single CMOS layer. Thus, many-PE cluster architectures greatly benefit from the increased routing resources tied to the vertical interconnects available in CMOS~2.0.  An initial study was presented in \cite{mempool3d} where a \gls{LOL} split was shown to improve interconnect latency, allowing single-cycle access to many memory banks for a large number of PEs, with major reductions in wirelength and power compared to an implementation in standard single-layer CMOS technology.

The above example shows that the most straightforward way to exploit three-dimensional integrations is the expansion of on-chip memories at the higher hierarchy levels, such as L1 or L2. However, CMOS~2.0 can enable a more radical approach, impacting also L0 memory, namely, larger register files, deeply embedded within the processor pipelines. Fine-grained partitioning in CMOS~2.0 can allow us to vertically integrate large L0 register files, which are essential resources for memory latency tolerance, commonly employed in CPUs, GPUs, and vector processors through register renaming, multithreading, and deeply-pipelined temporal execution, respectively. We envision CMOS~2.0 to enable significantly larger, vertically stacked L0 storage structures such as register files. These would not be implementable in standard CMOS due to the large area footprint and the resulting costly interconnects between the enlarged register files and the execution units.

CMOS~2.0 enables even more aggressive optimization. For instance, one could exploit the heterogeneity among different layers to implement, within a layer dedicated to L0 structures, special cells like multi-bit latches with dedicated routing resources. Using latches instead of SRAM has the advantage of enabling a much larger operating range at low voltage, hence reducing energy per access. Moreover, the area penalty with respect to standard SRAMs is more affordable thanks to the significantly increased area budget provided by the extra layer in the stack. We want to stress that this layer optimization approach can be generalized and used to optimize other logic building blocks, such as, for instance, the decoders and multiplexers needed for steering data from L0 storage to computing units. 

\gls{CIM} architectures are also poised to benefit from CMOS~2.0. \gls{CIM} has been shown to have great potential for improved efficiency in specific workloads such as dense linear algebra (which dominates deep neural network inference and training). \gls{CIM} merges computation with storage: it has been extensively explored in research \cite{cnm_cim}, and is getting increased traction in industrial products such as the Metis AIPU \cite{metisAIPU} from Axelera and EN100 \cite{EN100} from Encharge AI. \gls{CIM}'s main challenge is that memory structures enhanced with computing capabilities have significant density overhead compared to standard memories. Intuitively, by exploiting 3D stacking, this overhead can be significantly reduced. 
For example, \cite{3d_cfet_cim} proposes a monolithically stacked CFET-based SRAM array, significantly reducing RC parasitics and achieving high energy efficiencies. Additionally, the peripheral circuitry, consisting of data read/write and reduction computations, is implemented in FinFET technology for better latency without impacting the high-density footprint of the SRAM array.

\subsection{Novel architectural paradigms}
In addition to architecture and implementation enhancements that can be achieved by leveraging CMOS~2.0, the full exploitation of aggressive and heterogeneous three-dimensional integration can enable novel architectural paradigms, addressing the current dichotomy between flexibility and performance (and efficiency) in computer architectures. 

As an example, let us consider hardwired accelerators such as TPUs, featuring massive arrays of multiply-accumulate units that execute a single instruction with hardwired dataflow patterns, achieving high performance and energy efficiencies for compute-intensive workloads such as convolutions and matrix multiplications. CMOS~2.0 allows new degrees of freedom in the design of dedicated systolic computing structures. Through the integration of vertical interconnects, CMOS~2.0 enables high-bandwidth memory interfaces, allowing for dedicated ports for each PE to the memory, accelerating memory-intensive workloads previously not possible in 2D technologies.
Additionally, fast read or write-back of results from PEs to the memory can be achieved for compute-intensive workloads, avoiding larger latency penalties in hardwired systolic arrays. In other words, we can have much more "internal storage" in a systolic array (using L0-optimized storage layers), and, at the same time, much larger bandwidth and capacity for the L1 buffers needed to support input and output for the systolic structures. 

Early examples of exploitation of three-dimensional integration in the design of systolic arrays are presented in \cite{10929973, 8598454}. 
These works propose subarray partitioning of systolic arrays in the planar and vertical dimensions to reduce latencies in scaled-up array configurations. Subarray partitions in the planar dimension require high fanout physical wiring, affordable in 3D due to increased routing resources. Further, by exploiting the vertical dimension for reductions, latencies are further reduced, achieving higher PE utilization.

Another direction targets the need for supporting multiple data-parallel workloads with high efficiency in hardwired accelerators. A notable example is supporting low arithmetic intensity operators (e.g. matrix-vector multiplication) on systolic arrays primarily designed for matrix-matrix multiplication.  This dual-mode operation requires re-configurable interconnects and fine-grained power management of PEs, which is very expensive (in area and power) in traditional CMOS. CMOS~2.0 has the potential to enable architectural flexibility using specialized stacked layers. In the same direction, we could extend the capabilities of the processing elements in the systolic array. Moving from MAC units to more programmable logic requires local storage of control bits (microcode) for cycle-by-cycle operation sequencing. This approach has been tried many times in classical CMOS with \glspl{CGRA} \cite{CGRA_survey}. Unfortunately, \glspl{CGRA} have had limited commercial success because of their relatively lower efficiency compared to dedicated systolic arrays - in other words, the price paid for flexibility in \glspl{CGRA} has been too high in standard CMOS. 

CMOS~2.0 could enable a drastic reduction of 3D-\gls{CGRA} overhead, exploiting data-level parallelism, but also maintaining flexibility through reconfigurable vertical interconnects and abundant L0 storage previously not available in classic CMOS. 
As an example, a recent work \cite{10893709} reports improvements in wirelengths by placing SRAMs directly below configurable logic blocks (\acrlong{LOM}), achieving better timing and fewer insertion of buffers to fix long wiring delays otherwise present in 2D, with clear benefits on achievable frequency and energy efficiency. Additionally, the \glspl{TSV} around the less occupied bottom SRAM tier and dense vertical hybrid bonding power connections avoid power grid interruptions due to SRAM macros, reducing the observed IR drop.

Overall, the CMOS~2.0 paradigm offers exciting opportunities for novel architecture design. There is a need to transition from current-state-of-the-art 2-tier hybrid-bonded \gls{MOL} or \gls{LOL} splits to more fine-grained partitioning, at the level of L0 memories, and even combinational, sequential specialized layers. Partitions will be deeply embedded within the \gls{PE} pipelines. Further, going beyond von Neumann computing, \gls{CIM} architectures with high-density SRAM structures can be enabled with CMOS~2.0.
Finally, specialized architectures such as configurable systolic arrays and  \gls{CGRA}s, which are relatively inefficient in comparison with fully hardwired logic in traditional CMOS technologies, can become extremely competitive in CMOS~2.0. On the other hand, a few design implementation challenges are expected to rise in importance for CMOS~2.0, namely,  thermal and power management, as well as ensuring system reliability in the face of the increased impact of hard and soft faults. Section \ref{sec:reliability} will focus on ways to address these challenges.

\color{black}

\section{Electronic Design Automation}
\label{sec:EDA}

\color{color_eda}

From the beginning of VLSI, \gls{EDA} has been instrumental in handling the rapidly growing implementation complexity of digital systems. Specifically, logic synthesis has become extremely powerful to unlock intrinsic \gls{PPA} improvements with its automated mapping and optimization process. Fast-forward to today, \gls{EDA} is the cornerstone of every design and implementation flow process. In that role, it has to closely follow the emerging breakthroughs in process technology and extend its capabilities to prepare chip design for each successive generation. System scaling initially followed the momentum of geometry-centric scaling. Then, it evolved by further combining a diverse set of components into a System on Chip. And today, large-scale systems are composed of multiple chiplets through advanced packaging.

Products leveraging 2.5D and 3D integration \cite{amd_3D} \cite{amd_MI300} partition the system coarsely into different parts on the subsystem-level, and implement these in different pieces of silicon. The resulting chiplets then communicate through coarse pitch interfaces as indicated in Fig.~\ref{fig:technology:CMOS2.0vs3DIC}. 

In contrast, CMOS~2.0 leverages very fine pitch vertical interconnects to cut the connectivity graph of a system down to the abstraction layer of standard cells. In such a case, in contrast with conventional hierarchical design, the different tiers need to be implemented simultaneously to achieve the best QoR. Therefore, we claim that the CMOS~2.0 paradigm opens a new level of implementation complexity that goes far beyond the capabilities of existing 3D-IC toolsets.

Previous works tried to estimate the impact of fine pitch 3D integration on block-level for different schemes like \gls{MOL} and \gls{LOL} under various 3D stacking assumptions like \gls{F2F}/\gls{F2B} \gls{W2W}/\gls{D2W} hybrid bonding~\cite{macro_3d_stt} or monolithic integration~\cite{cascade2d}~\cite{m3dadtco}. In these, commercial 2D engines are tricked with modified technology files to represent the 3D arrangement of the design. While such approaches can elegantly evaluate some scenarios under certain conditions, they do not represent an expandable foundation for a holistic exploration platform as required for CMOS~2.0.

We distinguish between two steps that drive the requirements for implementation toolchains targeting CMOS~2.0 technology stacks:
\begin{enumerate}
    \item \label{EDA:requirements:support_3d} Introducing a multi-layer technology stack with very dense inter-layer connectivity
    \item \label{EDA:requirements:support_heterogeneity} Introducing heterogeneity and specializing the \gls{FEOL} and \gls{BEOL} of each layer for a specific functionality
\end{enumerate}
In the following, we discuss some of the major implications when approaching these challenges.

\subsection{Towards EDA for very fine pitch 3D stacks}

Assuming a system architecture designed for case \ref{EDA:requirements:support_3d}) and present in the form of its RTL, we believe the subsequent implementation toolchain requires significant extensions. First, when mapping Boolean expressions to gates, the synthesis engine will require the physical context, like the availability of placement and routing resources, as well as the inter-layer parasitics, to improve its QoR. Second, the optimization engines should support cell movements across layers to fully leverage the very fine inter-layer pitch. Third, physical awareness is required to cut the connectivity graph whenever it benefits over pure in-plane routing while satisfying the signal density limits dictated by the respective bonding technologies. With scaling of the bonding pitch, an increasing portion of the system will be able to harness this benefit. These examples highlight the increasing need for tightly coupling synthesis and placement engines to handle the finer interactions along the out-of-plane dimension. Further, the cost functions during placement, like global placement \cite{eplace_3d} or legalization \cite{legalizer_3d}, have to be redesigned. These engines typically leverage estimations of the routing engine; thus, it is crucial to embed characteristics of the complete stack into the parasitics model.
Eventually, the router needs adaptation to address critical details such as optimizing routes accessing cell pins now available on both the front-side and back-side of the active layers.

\subsection{Towards EDA for highly heterogeneous very fine pitch 3D stacks}

To expand on \ref{EDA:requirements:support_heterogeneity}), we now assume a system architecture designed for a particular CMOS~2.0 technology stack and again, in the form of RTL. In such a scenario, the engines used throughout the implementation process require even further extensions over the ones described for \ref{EDA:requirements:support_3d}). Mapping a certain Boolean expression to a particular cell of a particular technology is intrinsically assigning the z-component, or technology layer, to cells. As logical synthesis lacks the physical awareness that is crucial to drive educated mapping decisions, we see physical synthesis as a strict requirement in implementation flows targeting CMOS~2.0 technology stacks. Technically, cells in any \gls{FEOL} layer in the stack can be used to implement any functionality, but we believe that trying to solve the mapping problem leveraging that premise may be very challenging for large-scale systems due to the vast solution space. Instead, as every \gls{FEOL} and \gls{BEOL} layer in the stack has been designed purposefully to provide certain, potentially specialized functionalities to the system, we believe the definition of a function-based abstraction layer allows to guide the design and implementation flow.

\begin{figure}
    \centering
    \includegraphics[width=\linewidth]{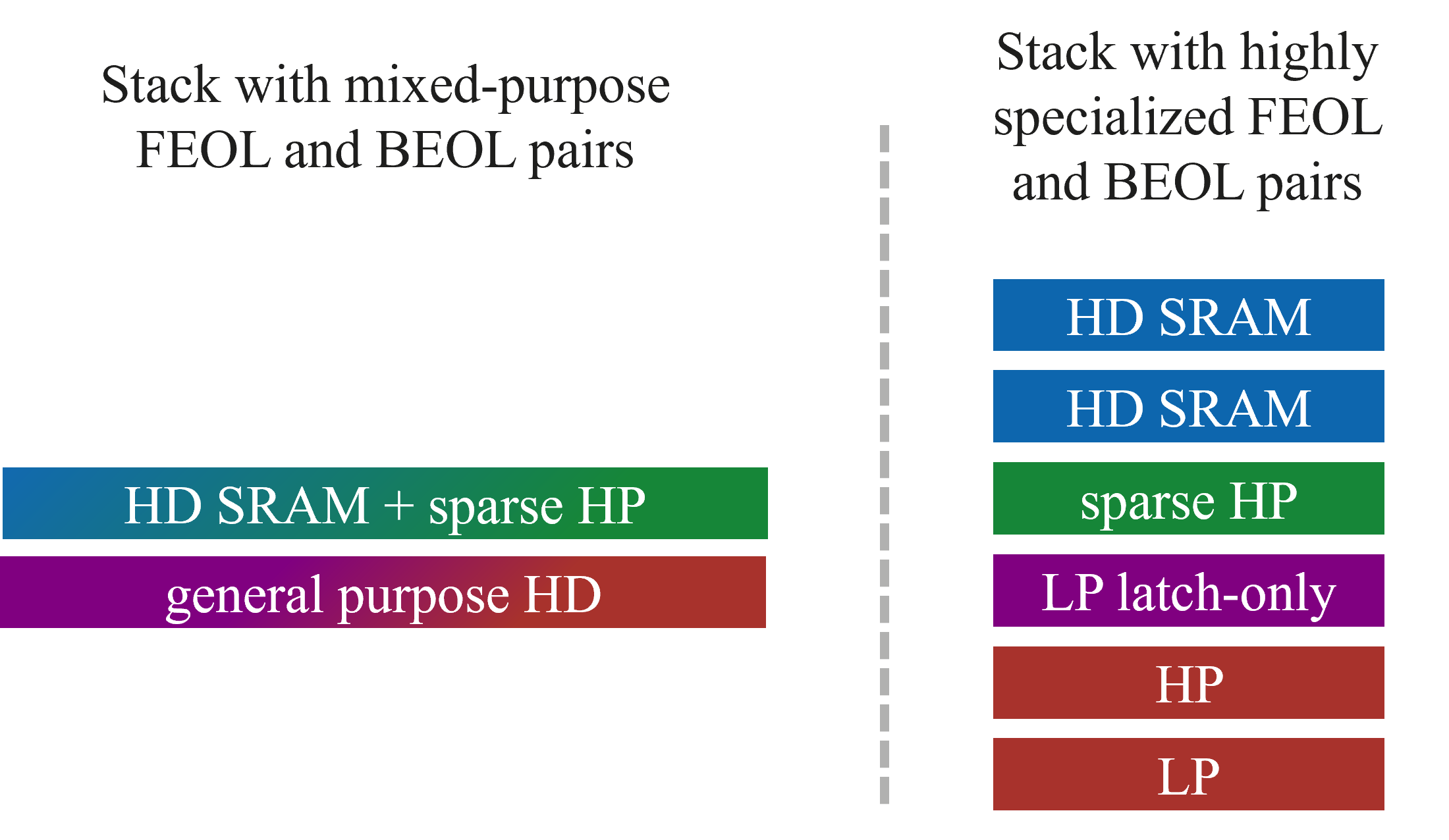}
    \caption{High-level examples of conceptual CMOS~2.0 stacks. Each rectangle represents a \gls{FEOL}-\gls{BEOL} pair and is labeled with its \gls{FEOL} optimization target. Further, each rectangle is annotated with a function through the function-color map introduced in Fig.~\ref{fig:arch_template}.}
    \label{fig:eda:function_annotated_stacks}
\end{figure}

Fig.~\ref{fig:eda:function_annotated_stacks} shows two very high-level examples of conceptual CMOS~2.0 stacks, where each box represents a \gls{FEOL}-\gls{BEOL} pair. Boxes are labeled with their respective optimization targets, such as \gls{HP} and \gls{LP}, at a high functional density unless stated otherwise. Fig.~\ref{fig:arch_template} shows a block diagram of a system architecture template with different layers of abstraction. It further defines a function-color map used for the boxes in Fig.~\ref{fig:eda:function_annotated_stacks}, too. By considering the colors that have been assigned to different blocks of the system in Fig.~\ref{fig:arch_template} as functional hints, it may be possible to link these hints to layers in a technology stack that has been annotated like in Fig.~\ref{fig:eda:function_annotated_stacks}.

This abstraction could even be utilized during the design flow when targeting a CMOS~2.0 platform. Early system architecture design benefits from a quick feedback loop to match design decisions with technology trends. Since CMOS~2.0 requires respecting the utilization across the different active tiers, linking architectural elements to particular layers through their functionality allows a first-order 3D-floorplan estimation and steers further design decisions. In the context of physically aware synthesis and placement, this mapping can be used to yield an early instance-level layer assignment, which can be refined by the different engines in the subsequent steps. While the functional hints in case of Fig.~\ref{fig:arch_template} may closely follow the logical design hierarchy, the granularity of hints is not tied to it, and these could theoretically be applied at the instance level. However, the logic hierarchy lends itself to hint-annotation by the designer. Future design and implementation flows could incorporate the derivation of functional hints as part of the \gls{EDA} toolchain.

In contrast to \ref{EDA:requirements:support_3d}), the refinement of the initial layer assignment during synthesis in \ref{EDA:requirements:support_heterogeneity}) requires a redesign of the underlying mapping cost functions. For example, using silicon area as a proxy for mapping complexity is insufficient to judge the quality of a mapping on its own. Certain layers could have been designed to purposefully exhibit a smaller transistor density to cater to a particular purpose, thus requiring a holistic stack view to also incorporate the implications on cross-layer utilization and connectivity.

Similarly, the cost functions of the placement engine require additional terms to capture the heterogeneous technology characteristics. As the effective timing cost of an interconnect depends on the \gls{FEOL} technology of the driving cell and the wire segments in potentially multiple \glspl{BEOL}, pure \gls{HPWL} is not sufficient to capture the characteristics of highly heterogeneous CMOS 2.0 technology stacks. Instead, connections could be partitioned into segments that model in-plane and out-of-plane connectivity per \gls{BEOL} that is traversed. Based on these, it may be possible to estimate a first-order interconnect cost that can reflect the highly heterogeneous \gls{BEOL} characteristics.

Further, as the \gls{FEOL} technologies in the stack may exhibit very heterogeneous drive currents, moving the driving cell of a net from one tier to another can lead to a discontinuity in drive capability and thus interconnect timing cost. Similarly, input pin capacitances can change abruptly when reassigning cells to different tiers. Therefore, the engines require the holistic modeling of both the interconnects and the \gls{FEOL} technology of each endpoint.

Block-level implementations assemble a design using standard cells. Thus, the partitioning schemes proposed in Fig.~\ref{fig:technology:CMOS2.0partition} require special consideration to take full effect. With \gls{FEOL}-\gls{BEOL} pairs dedicatedly optimized towards a very particular functionality, the layer assignment should be aware of the function-driven partitioning schemes constituting the stack.
For example, combinational/sequential partitioning allows for concentrating the clock-tree distribution to particular layers in the stack. Moving sequential cells into the combinational-optimized layer would reduce leverage of the functional specialization and necessitate distributing the clock signal in an additional layer.
Similarly, the \gls{AuC} partitioning of Fig.~\ref{fig:technology:CMOS2.0partition} necessitates information-storing circuits to be allocated within a certain structural context to benefit from the \gls{FEOL}-\gls{BEOL} specialization.

For CMOS~2.0, implementation tools must adapt and broaden their perception of technology. The CMOS~2.0 paradigm tries to cater to the architecture needs with the best technology choice available. Consequently, the \gls{EDA} implementation toolchain needs to be mindful to manage the highly heterogeneous requirements at all abstraction levels, and find smart solutions fulfilling them. \gls{EDA} has been key in leveraging technology solutions since its existence, and this time is no different.
\color{black}
\section{Reliability}
\label{sec:reliability}

\color{color_reliability}

The semiconductor industry faces mounting challenges as devices and systems reach unprecedented levels of complexity. Increases in transistor density and reductions in feature size enable significant performance gains, but also intensify manufacturing variability and heighten sensitivity to runtime- and workload-induced fluctuations. Higher integration levels further raise current and power densities, complicating voltage regulation and thermal management.

While compute performance and memory bandwidth remain the primary objectives of technology scaling—and stand to benefit greatly from CMOS~2.0 concepts—scaling also introduces collateral effects that cannot be ignored. These include reduced manufacturing yield, diminished testability, weakened fault tolerance, elevated power consumption, and aggravated thermal issues. Left unaddressed, these factors risk becoming the primary bottlenecks of future scaling. CMOS~2.0 approaches offer promising pathways to mitigate these fundamental reliability challenges in advanced technology nodes.

\subsection{Reliability Wall in Advanced Technology Nodes}

A major challenge of continued scaling is {\em test access}—the difficulty of accessing hundreds of billions of transistors through only a few hundred to a few thousand pins. This challenge is compounded in {\em 3D heterogeneous integration}~\cite{dhananjay2021monolithic}, where multiple wafers are stacked and access points are severely limited~\cite{marinissen2009testing}. As system complexity grows, minor defects under certain environmental conditions (e.g., supply voltage variations, thermal gradients, process corners) can lead to failures. This has increased reliance on {\em parametric testing} and {\em System Level Test} (SLT), both of which drive up test complexity and cost~\cite{appello2021system}.

Another critical concern is {\em manufacturing yield}. In conventional 3D stacking, the {\em Known Good Die (KGD)} approach—bonding only dies that pass wafer probe tests—helps maintain yield~\cite{knickerbocker20122}. However, in CMOS~2.0 stacking, such pre-selection is not feasible. Without novel solutions, yield can degrade sharply as more layers are bonded, threatening the economic viability of the technology.

Additionally, higher design densities push current and power densities to new extremes, demanding more robust solutions for voltage regulation and heat dissipation~\cite{hesheng2022_voltage_regulation}\cite{oprins2021_cooling_solutions_3d}. Complex interactions—spanning chip placement, system architecture, and hardware–software coordination—further increase the likelihood of physical failures. These factors make it difficult and costly to guarantee defect-free operation in the field, underscoring the need for {\em runtime sensing and monitoring} to predict, detect, and mitigate failures dynamically~\cite{landman2022applying}.

\subsection{Breaking Through the Reliability Wall with CMOS~2.0}

CMOS~2.0 offers new opportunities to address these scaling-induced reliability barriers in highly integrated systems. Its vertical connectivity, combined with {\em layer specialization}—segregating, for example, sequential vs. combinational logic or digital vs. analog functionality—opens novel avenues for resolving the {\em test access} bottleneck. For instance, consolidating all system bistables and flip-flops into a single layer would make it feasible to integrate additional hold latches into scan flip-flops within unused spaces. These latches improve delay-test coverage, enabling higher-quality testing and facilitating {\em state snapshotting} without disrupting normal operation—an advantage during SLT phases for diagnosis and debug.

CMOS~2.0 can also {\em boost yield} via process optimization per specialized layer. By tailoring manufacturing parameters to the functional requirements of each layer—e.g., high-drive logic, high-density logic, synchronous logic—the process can be simplified and refined, improving per-layer yield and, ultimately, the overall system yield.

For {\em defect tolerance}, especially in wafer-scale or other highly integrated systems, CMOS~2.0 enables solutions beyond memory-specific techniques such as {\em Error Correcting Codes} (ECC) or redundancy repair. By physically structuring all system bistables into a dedicated layer, they can be treated as an organized array—allowing memory-like repair strategies to be applied to flip-flops and latches. This architectural shift also mandates that {\em fault tolerance be integrated from the outset}, enabling systems to operate reliably even in the presence of defects. This has major impacts on how computing architectures will be designed in the era of CMOS~2.0.

\subsection{Integrating Silicon Lifecycle Management (SLM) in CMOS~2.0}

An increasingly important approach to reliability is \gls{SLM}—a holistic strategy spanning design, manufacturing, deployment, operation, maintenance, and end-of-life~\cite{tahoori2024silicon}. In the design stage, close collaboration between manufacturers and system architects ensures performance, power, and reliability goals are met while accounting for system-level constraints such as power consumption, thermal management, and security. Manufacturing focuses on process control, yield optimization, and quality assurance; deployment ensures seamless system integration; and operation emphasizes continuous monitoring, diagnostics, and maintenance.

In CMOS~2.0, \gls{SLM} can be deeply embedded into the chip’s infrastructure. Dedicated layers—optimized using specialized \gls{FEOL} processes—can host dense arrays of analog sensors, global voltage regulation, test structures, and system control logic without being constrained by the scaling requirements of compute layers. These sensors enable {\em fine-grained, real-time monitoring} and rapid feedback for adaptive system management, while integration with {\em Built-in Self-Test} (BIST) and {\em Design-for-Test} (DfT) layers supports on-chip parametric measurements. This separation of infrastructure from compute elements minimizes overhead and makes large-scale \gls{SLM} deployment economically viable.

In summary, the CMOS~2.0 concept can be leveraged to fundamentally overcome the reliability wall in advanced technology nodes and in highly integrated, complex systems. Its implications extend beyond defect detection and repair to the architectural level, reshaping future computing systems to be inherently fault-tolerant and lifecycle-aware.

\color{black}
\section{Conclusion}
\label{sec:conclusion}
Ultimately, CMOS 2.0 is a major disruption for the industry and requires rethinking the entire ecosystem from system architecture design down to technology definition. Such a revolution can only happen with a cross-disciplinary approach as many well-established assumptions may break and basic principles will most likely have to be revisited. At the same time, CMOS 2.0 must capitalize on all the progress that has been made in the last decades in VLSI design. However, as grandiose as the challenges are, so are the potentials to maintain an industry on a so far successful scaling path.
\bibliographystyle{IEEEtran} \bibliography{tex/ref}
\end{document}